St Edmund's College

# Georges Lemaître and the foundations of Big Bang cosmology

## Simon A. Mitton

Georges Lemaître was a remarkable contributor to the advancement of cosmology in the heady years following the two great revolutions in theoretical physics in the last century: general relativity and quantum mechanics. In the past two decades the impact of his key publications has advanced from very little to enormous. Among other landmark achievements, Lemaître was the first cosmologist to estimate the rate of expansion of the Universe, later known as the Hubble constant. In 2018 the International Astronomical Union renamed the Hubble law as the Hubble–Lemaître law in recognition of his foundational contribution to the theory of the expanding Universe. This biographical review examines: factors that influenced his approach to the philosophy of science, particularly cosmology; his engagement with the network of contemporary cosmologists; and the circumstances of his withdrawal from research in cosmology and the further promotion of 'the fireworks universe' after the early 1930s.

**Author:** Simon A. Mitton, *sam11@cam.ac.uk*.

St Edmund's College, Cambridge CB3 0BN, United Kingdom

# Georges Lemaître and the foundations of Big Bang cosmology

## Simon A. Mitton

Georges Lemaître was a remarkable contributor to the advancement of cosmology in the heady years following the two great revolutions in theoretical physics in the last century: general relativity and quantum mechanics. In the past two decades the impact of his key publications has advanced from very little to enormous. Among other landmark achievements, Lemaître was the first cosmologist to estimate the rate of expansion of the Universe, later known as the Hubble constant. In 2018 the International Astronomical Union renamed the Hubble law as the Hubble–Lemaître law in recognition of his foundational contribution to the theory of the expanding Universe. This biographical review examines: factors that influenced his approach to the philosophy of science, particularly cosmology; his engagement with the network of contemporary cosmologists; and the circumstances of his withdrawal from research in cosmology and the further promotion of 'the fireworks universe' after the early 1930s.

## 1. Recognition by the International Astronomical Union 2018

Georges Henri Joseph Édouard Lemaître (1894–1966) died three days after learning from his successor at the Catholic University of Louvain, the Belgian mathematician Odon Godart (1913–96), that Arno Penzias and Robert Wilson had discovered the cosmic microwave background radiation. Despite being gravely ill with cardiac failure and leukaemia, Lemaître lucidly thanked his colleague for telling him of this finding, which confirmed the explosive origin of our Universe, just as Lemaître had suggested in 1931.

More than five decades since his passing, Lemaître's reputation has risen remarkably, from near obscurity to widespread acclaim as the 'father of the Big Bang'.[1] At the Thirtieth General Assembly of the International Astronomical Union (Vienna, Austria, 2018 October 20–31) a lively discussion was initiated by the promulgation of Resolution B4 'on a suggested renaming of the Hubble Law'. This led by turns to the recommendation that 'from now on the expansion of the universe be referred to as the Hubble–Lemaître law'. The purpose of the resolution was 'to honour the intellectual integrity of Georges Lemaître that made him value more the progress of science rather than his own visibility'. This review explores how Lemaître's research from 1927 to 1946 contributed to our present picture of Big Bang cosmology.

## 2. Early life and education

Georges Lemaître had a fascinating and varied life in science. As a Catholic priest he made an intriguing academic, working in both heavenly theology and cosmical theory. During Lemaître's lifetime his older contemporary, Edwin Hubble (1889–1953), was idolized in the United States for discovering that the galaxies are receding, while in Britain Fred Hoyle (1915–2001) entertained the public and outraged professionals with his Steady State model of the Universe.[2] Lemaître knew both Hubble and Hoyle and kept on amicable terms with them.

From the late 1930s until the early 1960s, observational data on the distant Universe was so sparse that theorists were largely free to pick and choose what numerical values to use for the parameters in their equations. Throughout this period, what we now term observational cosmology was mostly carried out by Hubble at Mount Wilson with the 100-inch (2.5-m) telescope until 1949 July, when he suffered a heart attack while on vacation. Thereafter, his protégé Allan Sandage (1926–2010) took over the observational cosmology programme with the Palomar 200-inch (5-m), which had been specifically designed for research on the most distant galaxies.

Sandage undertook a quest to determine two numbers of deep cosmological interest: the rate at which our Universe expands, and the degree of acceleration in that rate. Lemaître, as well as other theorists, took the keenest interest in the data that had been acquired by Hubble and his successors because the rate of expansion and its acceleration are the only observables that could be used to differentiate between competing theories of the origin and nature of the Universe.[3]

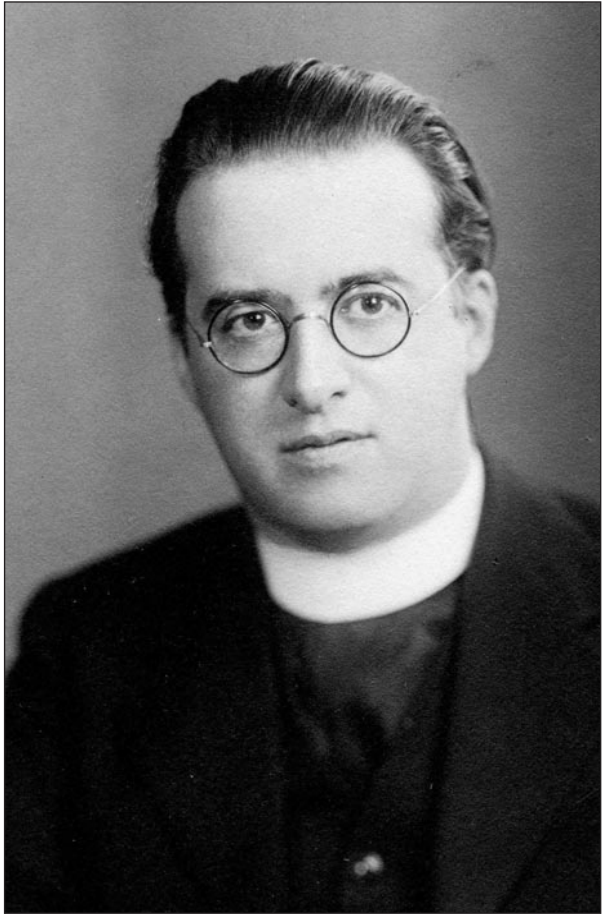

*Fig 1: Georges Lemaître in the early 1930s when he was at the height of his fame, a few years after publishing his groundbreaking paper on the expansion of the Universe. The English-language version of his 1927 paper appeared in 1931. (Archives Louvain)*

### 2.1. *Family and upbringing*

Who was Georges Lemaître and what aspects of his upbringing influenced his later career? His ancestry can be traced through four generations: his great-great grandfather Clément enlisted in Napoleon's army and fought at Waterloo. The devout Catholic family became strongly patriotic after the Belgian Revolution of 1830, which had led to the 1839 Treaty of London that recognized Belgium as an independent and neutral country.

The Lemaîtres hailed from Charleroi, a former garrison town, positioned strategically to safeguard the roads to Antwerp and Brussels from the marauding French. By the mid-nineteenth century it had become a thriving, heavily polluted, industrial city with an economy based on coal mining, iron foundries, glassworks, and tobacco factories. Growth and trade accelerated after the canalization of the Sambre river in 1828.

Sixty years later, when Joseph-Achille Lemaître (1867−1942), the father of Georges, had graduated from the Law School of the University of Louvain in 1889, his proud father set him up in business by gifting him a quarry near Antwerp and glassworks close to Charleroi. Joseph and his wife Marguerite's three children (a fourth died in infancy) were raised to respect the tenets of the ruling industrial class. These included personal dignity, professional integrity, devout commitment to the Roman Catholic faith, and loyalty to the established institutions, civil, societal, and religious.[4]

Georges, the eldest of the children, was born on 1894 August 17. He grew up in Belgium's *pays noir,* a dark landscape disfigured by huge slag heaps and blast furnaces stretching west to the horizon. His father, an innovative industrialist, worked to improve the family glassworks business, so we can imagine that from an early age Georges became familiar with fiery furnaces and their glowing ashes.

At the age of 10 he commenced a classical education at the Jesuit Collège of Sacré-Coeur, noted for its emphasis on academic excellence. There he received a thorough grounding in Greek and Latin, including verse and rhetoric.[5] By his second year his outstanding talent at mathematics had become evident.

Meanwhile his father was experimenting with new technique for working glass and was on the point of a technical breakthrough when a major fire destroyed his factory. The Lemaître family faced ruin. When he sadly surveyed the devastation, Joseph's top priority was to settle with his debtors and pay the workforce, which he did by borrowing from the wider family. Then he secured a senior management position dealing with commercial loans at Société Générale in Brussels, and the family resided in an imposing townhouse.

Georges, by then age 16, was enrolled at the nearby Collège St-Michel, a Jesuit school opened in 1905, with extensive teaching laboratories and an enormous chapel. In his final year at the Collège, Georges studied advanced mathematics and physical sciences while he prepared for admission to the School of Engineering at the Catholic University of Louvain. A future career as a highly paid mining engineer beckoned, and he would be able to rebuild the family's finances. Or so it must have seemed.

### 2.2. *A change of direction*

Georges began his foundation course in engineering in 1911 September and also signed up for a diploma course in philosophy. His interest in the history of mathematics began about that time. He read Euclid in Greek and the works of Euler, Gauss, and Jacobi in Latin. By the end of his final year he had distinguished himself in mathematics and physics but not in engineering. Was he beginning to doubt his calling as a mining engineer?

The prosperous economy of Belgium was critically dependent on the coal mines of Wallonia in southern Belgium and the enormous production from the copper mines in the Belgian Congo, where Jesuits were busy with establishing schools. But Georges had also developed a passionate interest in philosophy.

He was attracted spiritually to the Roman Catholic priesthood but he dithered over whether to join a religious order noted for its scholarship, such as the Jesuits, or become a diocesan priest, which might leave him with a little spare time to do advanced mathematics. He was still mulling this over when the decision was taken out of his hands by the commencement of the Great War on 1914 August 4.

### 3. Patriotic service on the Western Front

The government of neutral Belgium refused the German Army safe passage to France, which compelled Britain, as a guarantor of Belgium's neutrality, to enter the war.[6] Georges and his younger brother Jacques (1896–1967), both deeply loyal to king and country, immediately enlisted. Georges was despatched to the front line, inadequately armed with a French M1874 Gras service rifle that had no magazine. It fired a single shot but it was an improvement on the Charleville musket that it had replaced in 1874.

The well-drilled Wehrmacht attempted to occupy the whole of Belgium in order to create a corridor through which they could invade northern France. Their troops flooded in and, by mid-October, they had the Belgian army bottled up in the northwest along the river Yser. Georges experienced plenty of action during the fiercely fought battle of October 16–30, with Belgium suffering 3,500 losses and 15,000 wounded.

The tide of battle changed dramatically when the Belgians opened the sluice gates at Nieuwpoort to inundate the polders, forcing the Wehrmacht to abandon its attempt to capture the front line. Thereafter the Belgian Army dug in on a section of Western Front that they held until 1918. Importantly, their soldiers secured Nieuwpoort, which would become the entry point for relief supplies from America. Had they failed to do so, Belgium would quite possibly have suffered mass starvation by 1917.

It was a horrible four years for the infantry, living and sleeping in insanitary trenches. Belgium lost another 7,000 troops to typhus. Although Georges was transferred from trench warfare to land warfare on 1915 July 3, the living conditions remained miserable. However, from that date on he had time to keep up his interest in science and delighted his comrades with stories of discovery.

Astonishingly, he not only mastered Poincaré's *Électricité et Optique*, he also diligently recorded the positions of his artillery battery each day on blank pages in the book. During his introductory course on ballistics, the jovial scholar politely pointed to a trigonometrical error in the official artillery manual. This resulted in a charge of insubordination from which his army career never recovered: his promotion to second lieutenant was blocked on the grounds of bad character. Nevertheless,

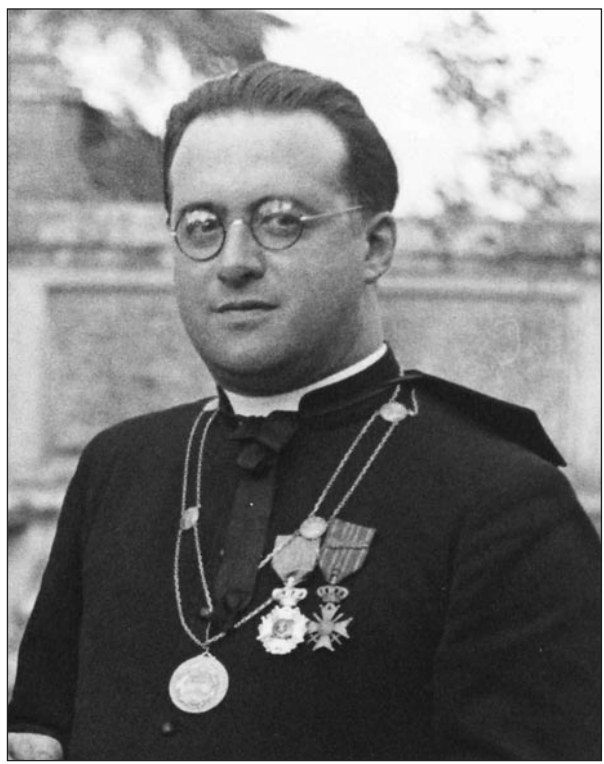

*Fig 2: Georges Lemaître attired with his war medals in a characteristically jovial pose at a regimental reunion, possibly late 1940s. On the right is his highly distinguished Croix de Guerre. (Archives Louvain)*

in peacetime he merrily mingled in many reunions of his regiment, proudly sporting his Croix de Guerre with Palms (Figure 2). He was one of only five front-line troops to be decorated by the Commander-in-Chief, King Albert I.

Attired in his shabby squaddie uniform, Lemaître resumed his studies at Louvain on 1919 January 21.[7] He dumped mining engineering and switched to an indepth study of higher mathematics and physics, concentrating on the spectacular advances in electromagnetism and relativity that had been made in the previous halfcentury. In 1920 July, Louvain conferred on him its doctorate in science (DSc), in effect his diploma to teach those subjects in higher education. His discharge from the army came through on 1919 August 19.

### 4. Three years of discovery

Having served under strict military orders for five years, Lemaître concluded that the enclosed life of a religious order was not for him. Instead he was fast-tracked for ordination as a secular cleric at the liberal Grand Séminaire de Malines. Here his intelligence quickly made the right impression on his superior, Cardinal Désiré-Joseph Mercier (1851–1926), Primate of Belgium, who encouraged the talented seminarian not only to keep up with the latest developments in general relativity but also to develop his interest in philosophy.[8]

In his encouragement of Lemaître's scientific career, the Cardinal's timing was perfect. On 1919 May 29,

Arthur Stanley Eddington (1882–1944), Plumian Professor of Astronomy and Experimental Philosophy at the University of Cambridge, had photographed the total eclipse of the Sun from a coconut plantation on the island of Principe, 220 km off the west coast of equatorial Africa.

On November 6 Eddington had announced to a specially convened meeting at the Royal Society that measurements of the positions of stars on his eclipse plates had confirmed the second classical test of Einstein's general theory of relativity, the extent to which starlight is bent by the Sun's gravitational field. Overnight, Eddington had soared to worldwide fame, and a publishing boom of books and papers on general relativity followed.

After Lemaître had mastered the technical details of general relativity, he spent 1921–22 writing a 131-page dissertation on Einstein's physics.[9] He submitted this review with his successful application to the Belgian Ministry of Arts and Sciences for a fellowship which enabled him to travel overseas for three years, extending his studies of general relativity in England and earning a doctorate in the United States. In just three years he had not only completed the seminary training but had also acquired a thorough grasp of the most fundamental aspects of the theory.

Mercier ordained Lemaître on 1923 September 22. Ten days later he alighted at the Port of Dover on the first stage of an exceptional voyage of academic discovery that would last three years.

### 4.1. *With Eddington in Cambridge*

Lemaître spent the first of these years in Cambridge, settling into fundamental research under the donnish tutelage of Arthur Eddington, who was surprised by Lemaître's excellent knowledge of the intricacies of solving the field equations of general relativity. At Cambridge Observatory, the highly decorated Belgian war veteran would have been somewhat in awe at the prospect of working with England's most famous pacifist scientist on the nature of the Universe.[10] Together they took up the challenge of discovering what, if anything, Einstein's field equations could reveal about the physics of our Universe.

Lemaître's period of study in Cambridge was highly significant for his future career in cosmology because Eddington was among the most distinguished astrophysicists in the world. When Lemaître arrived on the scene, the second edition of Eddington's great text on general relativity was in press,[11] and he was hard at work on another advanced monograph on the physics of stellar interiors.[12]

It was thanks to Eddington that Lemaître learned how to solve the differential equations of relativity by numerical methods. In Belgium, Lemaître had received no instruction on astrophysics whereas, at Cambridge, he attended Eddington's lectures on astrophysics, taking a keen interest in Eddington's discovery that the total radiation from a star (absolute luminosity) was determined by its mass.[13] Eddington had arrived at this conclusion (the mass–luminosity relationship) by considering the equilibrium of a stable star, in which the force of radiation pushing outwards is balanced by the gravitational force pulling inward.

There had already been attempts to solve the equations of general relativity when it was applied to the entire Universe. In 1917 Einstein himself had been the first to do so, finding a solution for an infinite static universe – devoid of motion – in which the attractive force of gravity on ordinary matter is counteracted by a mysterious cosmological constant.[14]

A different model had been proposed by the Dutch mathematician and astronomer Willem de Sitter (1872–1934), but his universe was devoid of matter. Neither solution approximated reality, but they did demonstrate the difficulty of finding a model that would fit the facts. The search for such a solution became the main focus of Lemaître's three years of study abroad.

During his Cambridge year he made an original contribution to the development of general relativity by producing a clearer mathematical understanding of what is meant by the concept of simultaneity in a four-dimensional universe with three spatial dimensions measured with a ruler, and a time dimension measured by a clock.[15] In a letter of recommendation dated 1924 December 24 addressed to Théophile de Donder (1872–1957) of the Université Libre de Bruxelles, Eddington included a warm tribute to his former student:

> I found M. Lemaître a very brilliant student, wonderfully quick and clear-sighted, and of great mathematical ability. He did some excellent work here, which I hope he will publish soon. I hope he will do well with Shapley at Harvard. In case his name is considered for any post in Belgium I would be able to give him my strongest recommendations.[16]

### 4.2. *The measure of the Universe*

Eddington introduced Lemaître to the interesting properties of Cepheid variable stars, particularly their application as 'standard candles' for the extragalactic distance scale. Lemaître learned that in 1912, at Harvard College Observatory (HCO), Henrietta Swan Leavitt (1868–1921) had established that the apparent magnitudes of 25 Cepheid-type variable stars in the Small Magellanic Cloud decreased almost linearly with the logarithms of their periods.[17]

Within a year, the Danish astronomer Ejnar Hertzsprung (1873–1967) was able to calibrate the relationship between a Cepheid's period and its luminosity, thus recognizing Cepheids as 'standard candles' for illuminating the extragalactic distance scale.[18] In 1914, while on the staff of Mount Wilson Observatory, Harlow Shapley (1885–1972) suggested that the periodicity of Cepheid luminosity arises because they are regularly pulsating.[19] By 1919 Eddington had a well-developed thermodynamic model to account for the phenomenon of stellar pulsation, which he applied to Cepheids in a successful test of his mass–luminosity relation.[20]

When Shapley had been appointed the director of HCO in 1921, he was already the world expert on using periodic variable stars (RR Lyrae stars) to gauge distances in our Galaxy. He made a big push to improve the calibration of the period–luminosity relation so that it could be used with confidence to establish the extragalactic distance scale. This intensive study recognized that there is more than one kind of Cepheid variability and he showed how to recognize their differences.[21]

Shapley undoubtedly laid the foundations for Edwin Hubble's estimates of the distances to spiral nebulae (as galaxies were then called).[22] In the observing season of 1923–24 Hubble was busy observing the periods of 22 Cepheids in M33 and 12 in M31 with the 60-inch (1.5-m) and 100-inch (2.5-m) telescopes at Mount Wilson.[23]

## 5. Canada and the United States 1924–25
In 1924 June Lemaître bade farewell to his new friends at the University of Cambridge and returned home for a short break, after which he took a transatlantic liner to Quebec and Montreal to commence an adventurous year of travels in North America. On 1924 August 3 he caught up with Eddington, who was about to give a major public lecture during the Toronto Meeting of the British Association (BA).[24] Eddington was by then Great Britain's most prestigious physicist, on hand in Toronto to promote British science with a presentation on how he had confirmed general relativity by observing the displacement of starlight during the 1919 eclipse.

Lemaître's lasting impression from this conference was meeting Ludwick Silberstein (1872–1948), a Polish–American physicist who had published an acclaimed textbook on relativity in 1912. At the University of Toronto, Silberstein promoted special and general relativity as new staples of the physics curriculum. He had shown that, in de Sitter's universe of negligible mass, the light of a distant object will be redshifted by an amount dependent on its distance and he claimed that his formula was underpinned by observations of globular clusters.[25]

Establishment astronomers dismissed the idea of a link between redshift and distance, but young Lemaître reacted favourably. Although the mathematician was still a novice when it came to observational astronomy, he became rather animated on hearing first-hand of a possible link between two observables – velocity and distance – and general relativity.[26] Lemaître became anxious to find out more from expert observers about how variable stars could be used to fathom the extragalactic Universe.

A five-hour train ride took him north-west of Toronto, past Lake Ontario, to Ottawa, where he spent four weeks at the Dominion Observatory. He met up with spectroscopist François C. P. Henroteau (1889–1951), a former astronomer of the Belgian Royal Observatory, who had had been obtaining Cepheid light-curves for several years. Henroteau's monumental assessment of the observational data then available led him to conclude:

> Shapley's period–luminosity relation should perhaps be regarded more or less as a curve of statistical averages … and that the true relationship should involve [temperature] as well. Whether, and by how much, this would affect conclusions as to the scale of the universe can scarcely be determined.[27]

### 5.1. *Working with Shapley at Harvard*
In 1924 September Lemaître crossed the border from Canada into the US, eagerly anticipating his year at Harvard University in the 'Other Cambridge' in Massachusetts, where he was about to learn more of Shapley's cosmic distance scale. In Cambridge Lemaître resided at 1 Cleveland Street, the rectory of St Paul's Church, hard by Harvard Square (today a private residence). On November 13 he assured Cardinal Mercier that he was fully participating in the religious life and liturgy of St Paul's Parish.[28]

Before he could commence his studies with Shapley, Lemaître needed to register as a graduate student working for a PhD, but he was unable to do that at Harvard University because the Observatory had no teaching faculty and it did not offer graduate courses. Instead Georges registered with the Massachusetts Institute of Technology (MIT), with the inscription 'Rev. Georges Lemaître. Belgian Fellow 1924–1925. D. Sc. Louvain July 1920.' MIT did not recognize the courses he had followed to obtain his DSc as sufficient training to proceed to a PhD, so he was required to repeat graduate courses in mathematics and physics that he had already passed at Louvain.

Under Shapley's direction, Lemaître improved the method for calculating the period of oscillation of a pulsating star with a known mass and spectral type. To do that he revised Eddington's analysis of Cepheids in 1919 that had verified the mass–luminosity relationship and he invented a graphical presentation of four properties of a pulsating star. The relationship between luminosity, period, mass, and temperature could be read quickly from this graph, so that if three of the properties were known the fourth could be estimated.[29]

The relevance of this result to his future research in extragalactic astronomy is that Lemaître's method employed well-established classical physics to account for the numerical value of the period in widely different types of stars: short-period variables, typical Cepheids, and long-period variables. Astronomers could therefore have confidence in Cepheids as standard candles because the cause of their pulsation had been clearly explained by the physics of radiative equilibrium.

### 5.2. *The distance of the Andromeda nebula*
With that contribution to astrophysics published, Lemaître worked on a way to fix the deficiencies of Einstein's static universe packed with mass and de Sitter's dynamic universe devoid of mass. He felt that de Sitter's empty universe was where to start, and his first contribution to theoretical cosmology appeared in 1925.

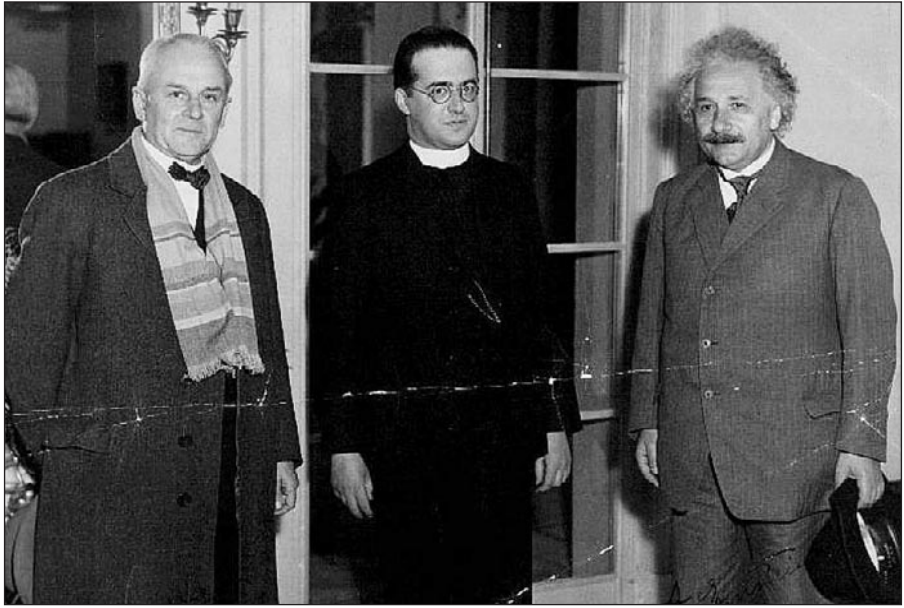

By taking a more sophisticated mathematical approach than de Sitter, Lemaître obtained a model describing a universe with a radius of curvature that depended on time. It seems this may be when he first became intrigued by the possibility of an expanding universe.[30] However, the problem with his model universe was that space extended to infinity, which Lemaître the philosopher simply could not accept, because of 'the impossibility of filling up an infinite space with matter which cannot but be finite'.[31]

In the first six months of 1925, Lemaître actively networked to integrate with the rather small North American community of extragalactic astronomers. January saw him at the 33rd Meeting of the American Astronomical Society, Washington, D.C. The big event at this gathering was a communication, read aloud by Henry Norris Russell (1877–1939), of Hubble's latest observations of 12 Cepheids in the Andromeda nebula (M31).

The M31 Cepheids showed precisely the same pulsation characteristics as those in the Milky Way and the Small Magellanic Cloud, so they were reliable indicators of the distance to M31: 285,000 parsecs (930,000 light years).[32] That result impressed Lemaître so greatly that he immediately decided to use his travel budget to improve his knowledge of astronomy by visiting several universities and observatories.

### 5.3. Reflections in British Columbia

Following an Easter break in the Canadian Rockies, Lemaître made his public debut as a cosmologist on 1925 April 25 by presenting his paper on the shortcomings of de Sitter's world model at a meeting of the American Physical Society in Washington, D.C.[33] Lemaître made his carefully researched tour of the West Coast in May and June 1925, commencing at the

Dominion Astrophysical Observatory, British Columbia, where the director John Stanley Plaskett (1865–1941) warmly welcomed him.

Plaskett has been largely neglected by historians, despite his being a highly accomplished designer of spectroscopes and telescopes, who custom-built the spectrograph for the 15-inch (0.38-m) refractor at Dominion Observatory, Ottawa. That was so successful that he began lobbying for the construction of a 72-inch (1.8-m) reflector. The Minister of the Interior of Canada approved the project in 1913 February. The disk was completed and shipped by the Saint Gobain Company near Charleroi, Belgium, late in 1914 July, only a few days before the German invasion. As Lemaître gazed at the 72-inch reflector, then the largest in the British Empire, he must surely have beamed with pride on being informed that its mirror (1,970 kg, 0.3 m thick) had been cast in his place of birth.

However, Lemaître had sought out Plaskett for more than a tour of the telescope. The Canadian astronomer, thirty years his senior, had enormous experience in spectroscopy and in determining the radial velocities of stars. He had recently started an extensive programme to obtain the radial velocities of five hundred highly luminous giant stars (types O and B) in order to determine galactic rotation and the structure of the Galaxy. Plaskett greatly enjoyed talking about his work in the convivial company of scientific colleagues, and Lemaître learned much from him about the dynamics of our Galaxy.

### 5.4. Lick Observatory

The next staging post on Lemaître's 'Cook's tour 1925' was the Lick Observatory, atop Mount Hamilton, about 20 km east of San Jose, California. Its director, William Wallace Campbell (1862–1938), delighted the visitor with two admirable attractions. These were the 36-inch (0.9-m) refractor for visual astronomy, which had been in use since 1888 when it was the world's largest before being edged out by the 40-inch (1-m) at Yerkes. The other was the 36-inch (0.9-m) Crossley reflector that dated from 1895.

Lemaître would have been well acquainted with the outstanding observations on galaxies that the astrophotography pioneer James Edward Keeler (1857–1900) had carried out with the Crossley reflector.[34] Keeler had established the importance of using large reflecting telescopes in mountain observatories for exploring the extragalactic Universe. He estimated that 120,000 spiral nebulae (i.e. galaxies) were within the grasp of the Crossley reflector.

Following Keeler's untimely death from a stroke in the summer of 1900, Lick Observatory commissioned, at great expense, a magnificent large folio volume showcasing seventy-one reproductions of his photographs.[35] Lemaître probably saw this volume for the first time in 1923 when working with Eddington, and Shapley must surely have pointed him to Harvard's copy. Keeler's Plate 1 is a superb image of the Andromeda nebula (Figure 4), much more detailed than the splendid example that the pioneer British astrophotographer Isaac Roberts (1829–1904) exhibited at the Royal Astronomical Society in 1886.[36]

### 5.5. Radioactivity and cosmic rays
On 1925 June 18, Lemaître made his final stopover in California, at Pasadena, where he met Robert Andrews Millikan (1868–1953), the first president of Caltech.

*Fig. 4: James Keeler's classic photograph of M31, the Andromeda spiral galaxy, taken with the 36-inch (0.9-m) Crossley reflector of Lick Observatory in 1899. (Publications of the Lick Observatory Volume VIII, 1908)*

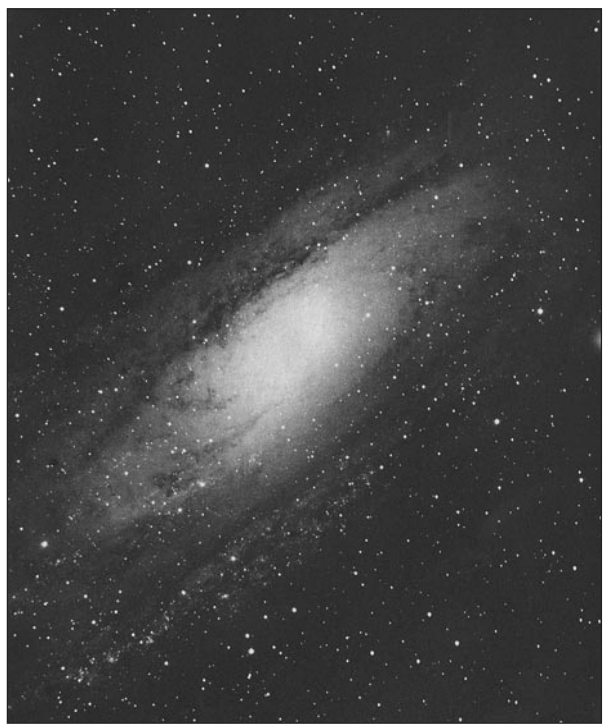

Millikan was in the midst of a decade-long project of research on cosmic rays, the nature of which was still a great mystery in the early 1920s.[37]

In the 1910s, physicists such as Ernest Rutherford (1871–1937) were researching 'radioactive emanations' by examining the radioactive properties of the Earth and its atmosphere.[38] Evidence was accumulating for the existence of a background of 'penetrating radiation' that could not be accounted for by emanation from the radium in their laboratories because it was ubiquitous in the open air.

Some physicists thought that uranium in the granitic rocks of Earth's crust was the likely source, but others were not so sure. It was in 1912, while conducting a total of seven dangerous and daring balloon flights at up to 5,350 metres (17,552 ft) altitude to measure the electrical conductivity of the atmosphere, that the Austrian-American physicist Victor Hess (1887–1964) discovered that this penetrating radiation was probably extraterrestrial.[39]

When Lemaître breezed through his office door, Millikan was within weeks of publicly confirming the extraterrestrial hypothesis.[40] Millikan, who coined the term 'cosmic rays',[41] had recently examined the evidence of the effects of the radiation on high-altitude lakes, concluding that they had found 'unambiguous evidence for the existence of [very high energy] rays of cosmic origin entering the Earth uniformly from all directions'.[42]

This encounter with Millikan would have provided Lemaître with vital first-hand knowledge of the novel field of cosmic ray physics that would come to play a key role in his construction of Big Bang cosmology, and in particular Millikan's hypothesis that protons and electrons had materialized from electromagnetic radiation, the primordial substance of the Universe.

## 6. Visiting Hubble and Slipher
Lemaître also called on Edwin Hubble at the Mount Wilson Observatory to see for himself the impressive 60-inch and 100-inch telescopes that Hubble and his night assistant Milton Humason (1891–1972) had employed to measure the extragalactic distance scale. As we have noted, Lemaître had already attended presentations by Hubble in Washington, D.C., but now at last he was conversing with Hubble in the dome of the 100-inch.

By this time, Hubble had Slipher's data on the radial velocities of spiral galaxies (which he used without one word of thanks) and his assistant Humason had commenced an attack on their distances using observations of their Cepheid variables. Hubble liked nothing more than putting on a good show for a distinguished visitor, although Prohibition meant there were no celebratory drinks.

Returning from the west coast via historic Route 66, Lemaître visited the Grand Canyon and Lowell Observatory, Flagstaff, where Slipher had been obtaining red-shifts for a dozen years. Lemaître knew that in 1922 February Slipher had sent Eddington the radial velocities of 41 spirals, almost all of which were unpublished. Slipher was no propagandist, so Eddington had done him a favour by reproducing the list in section 5 of *The Mathematical Theory of Relativity*, with the title 'Properties of de Sitter's spherical world'.

When it came to interpreting those velocities, Eddington became wary. Although he recognized that 'the great preponderance of positive (receding) velocities is very striking' the absence of velocities for spirals in the southern hemisphere meant the sample was biased, 'and forbids a final conclusion'. Nevertheless, he was the first theorist to understand that a more complete sample of extragalactic redshifts would open up the theory to observational constraints. And we can see that Lemaître's forensic questioning of the observers and his inspections of their facilities were positioning him to engage in cosmology.

Lemaître made one further foray, a second trip to Yerkes, where he attended an informal reunion of professional astronomers from the Chicago region. On that occasion he met a Cambridge postdoc, the mathematician Leslie John Comrie (1893–1950), who had been introducing scientific computational methods into student courses. This encounter introduced Lemaître to the great potential of large-scale mechanical computation for speeding up the reduction and the accuracy of astronomical data.[43]

# 7. The great breakthrough of 1927

Lemaître returned to Brussels and the family home on 1925 July 8. The following week he was off back to Cambridge for the Second General Assembly of the International Astronomical Union, from July 14 to 22. The Chancellor of the University of Cambridge, Earl Balfour (1853–1945), welcomed almost 300 visitors from 20 countries to a reception in the Senate House, following the conferment of honorary doctorates on William Campbell and Willem de Sitter in a colourful hour-long ceremony conducted entirely in Latin (as it still is today).

The following day, Eddington and his sister threw a garden party at the Observatory and, in the evening, Eddington's college, Trinity, served a magnificent conference banquet, accompanied by the finest claret and port from its cellars. It was the first time in ten months that Father Georges could enjoy a decent drink in convivial surroundings. He had opportunities to renew his acquaintance with Slipher, de Sitter, Hubble, and of course Eddington. For Lemaître, the IAU banquet was a momentous conclusion to his two years of travel and study with several distinguished cosmologists, all of whom welcomed him as an equal.

In 1926–27 Lemaître returned to MIT for the final year of his travelling fellowship. Back in Belgium in 1927 June he received news that his thesis had been approved without the need for an oral defence. And he was about to publish his great paper on the expanding Universe.

The University of Louvain promoted Georges Lemaître PhD to a professorship. From 1927 he taught classical mechanics to engineers and relativity theory to the mathematicians and physicists. His soaring breakthrough in cosmology had happened remarkably quickly. It was propelled by his deep knowledge of the extragalactic Universe.

He had taken great care to assess all of the observational data, to meet the observers and inspect their facilities, and to draw his own conclusions on their validity. Under Eddington's sure guidance he had become adept at the mathematical formalism of the general theory of relativity. He now put his learning and mathematical skills to good use by considering the strengths and weaknesses of the Einstein and de Sitter models.

## 7.1. *Friedman's expanding universe*

Einstein had dealt with the problem of the stability of a static universe by introducing a further factor into the field equations – the famous (or infamous) cosmological constant. This ad hoc fudge term had to be extremely small in order to preserve the agreement between general relativity and the observed planetary motions.

De Sitter likewise had introduced an adjustment, so that from 1917 the Einstein and de Sitter universes were each static and finite. Although de Sitter's universe had a redshift effect, that was a geometrical quirk of four-dimensional spacetime rather than an expansion of space. When Hubble announced a distance of 930,000 light years for M31, there was no observational evidence for expansion.

With hindsight we can see that the theorists uncovered the expansion first. Or, to be more precise, theoretical physicists discovered mathematical constructs that allowed a non-static universe.

In 1922, the Russian mathematician Alexander Friedman (1888–1925) started with a blank page.[44] He assumed that Einstein's fundamental equations were valid, by retaining the cosmological constant fudge factor, and he did not allow himself to be distracted by the contradictory nature of de Sitter's model.[45] He focused on models with positive curvature in the three spatial dimensions, and in 1924 considered models with negative curvature as well.[46]

When he allowed the radius of the curvature to vary with time, the equations sprang to life, with an infinite number of solutions in which the world can grow or shrink, according to the arithmetical sign of the cosmological constant. With the cosmological constant set to zero, the universe oscillated.[47] Friedman was the first to hit on non-static solutions that allowed the universe to expand, contract, collapse, and even to be born.[48] He commented on the impossibility of deciding what kind of universe ours is because of the inadequacy of the data then available.

In 1923, when producing an account of his findings for the public, Friedman mulled over the concepts of the origin of the Universe, cyclical universes, and a Big Bang, without connecting those thoughts to physics or astronomy. He suspected such notions were mere curiosities.[49]

Einstein issued a public rebuke of Friedman's solutions, dismissing them as incompatible with the field equations.[50] Eight months later he promulgated a humble retraction, agreeing that dynamic solutions were admissible.[51] All the same, it seems likely that Einstein thought of Friedman's solutions as a gimmick without cosmological merit. Friedman died of typhoid fever on 1925 September 16, about the time that Lemaître's professional career commenced.

### 7.2. *Betwixt Einstein and de Sitter*
In 1927 Lemaître's attention returned to the Einstein and de Sitter models that he had first discussed two years earlier (Section 5.2).[52] By this stage he had convinced himself that he must seek an intermediate solution that combined the merits of both models while eliminating their deficiencies. To handle the complexity of the analysis he made use his own choice of coordinate system, which he had unveiled in 1925 in Washington, D.C., at his cosmological debut. He was influenced in part by Silberstein's approach, which he had encountered first-hand in Toronto.

By 1927 Lemaître was convinced that our Universe is expanding, so at the very least he wanted to test if the equations had a solution with expansion. Following Einstein, he added the cosmological constant for good measure. Lemaître and Eddington considered all along that it played an essential role in the history of the Universe.

He progressed rapidly, soon hitting on a model in which the radius of curvature increased with time. It gave a natural explanation of Slipher's redshifts: that space itself was expanding and increasing the distances between galaxies. The redshift phenomenon occurs because the wavelength of a photon emitted from a galaxy increases in proportion to the increase in the size of the Universe during the time interval between emission to detection by an observer. This intuition took him straight to the velocity–distance relationship, formerly lauded as the Hubble law but now termed the Hubble–Lemaître law following the IAU vote in 2018.

Lemaître also calculated a value for the Hubble constant, the rate at which the Universe expands in kilometres per second (km/s) when the distance increases by one megaparsec (Mpc). Lemaître's value, 625 km/s/Mpc, was close to Hubble's value of ~500 km/s/Mpc. Lemaître was now convinced of the reality of the expanding Universe but he accepted that much improved data would be required to convince others.

### 7.3. *Dismissed as 'an abomination'*
Lemaître published his findings in 1927 in the Annals of the Scientific Society of Brussels.[53] That took a great deal of courage on his part because Einstein and his followers were still firmly attached to the static solution. In terms of the formalism, Lemaître's solution was a symmetrical spherical space that grew exponentially over time. It was identical to Einstein's solution in the infinite past and it would tend to a de Sitter universe in the infinite future.

Abbé Georges Lemaître, who invariably donned clerical dress, was quite at ease with these two infinities because, like Thomas Aquinas, he regarded faith (religion) and reason (science) as separate roads to the truth. At this time he was blissfully unaware of Friedman's solution. And Lemaître's brilliant 1927 paper met the same fate as Friedman's: the silent treatment, overlooked by the community, despite being published in a reputable journal with an international circulation.

In 1927 October Einstein participated in the 5th Solvay Congress in Brussels, where the main topic was quantum theory. Lemaître was not invited but, with Louvain being only 20 km from the capital, he opportunistically managed to catch Einstein's attention during a break. The two took a stroll in the Parc Léopold during which Einstein commented favourably on Lemaître's mathematical competence, although he rejected the notion of an expanding Universe as an abomination.[54] And on this encounter it was Einstein who directed Lemaître's attention to Friedman's work.

Lemaître found the news that Friedman had already made the breakthrough deeply unsettling but his confidence in Slipher's data encouraged him to continue to work on expansion as the key to interpreting the data on 'nebular velocities'. Nevertheless, Lemaître did not seek endorsements from Eddington or de Sitter. When Lemaître attended the IAU 1928 General Assembly in Leiden he must have engaged with de Sitter and Eddington, but there is no hint of this in the historical record.

## 8. Promoting a brilliant solution
Meanwhile, in 1925, Hubble had initiated a three-year project to estimate the distances of 24 galaxies for which velocities were already known. For the nearest galaxies, seven Cepheids were used as distance indicators, for the next thirteen galaxies the brightness of their most luminous stars was used, and for the four most distant galaxies he used the brightness of gaseous nebulae. It was a sparse sample, but it sufficed for Hubble to find the linear relationship between the velocity of galaxies and their distance that would eventually provide definitive evidence for the expansion of the Universe prefigured in Lemaître's neglected paper of 1927.[55]

Hubble risked his reputation with his 1929 paper because of the low quality of the data. He felt that his analysis 'supports validity of the velocity-distance relation in a very evident matter', and he gave a value of 500 km/s/Mpc for the 'distance effect' of the rate of increase of radial velocity. Einstein, when confronted with Hubble's results, which were much better than those available to Lemaître two years earlier, accepted the expansion of the Universe, conceding that there was no need for a cosmological constant.

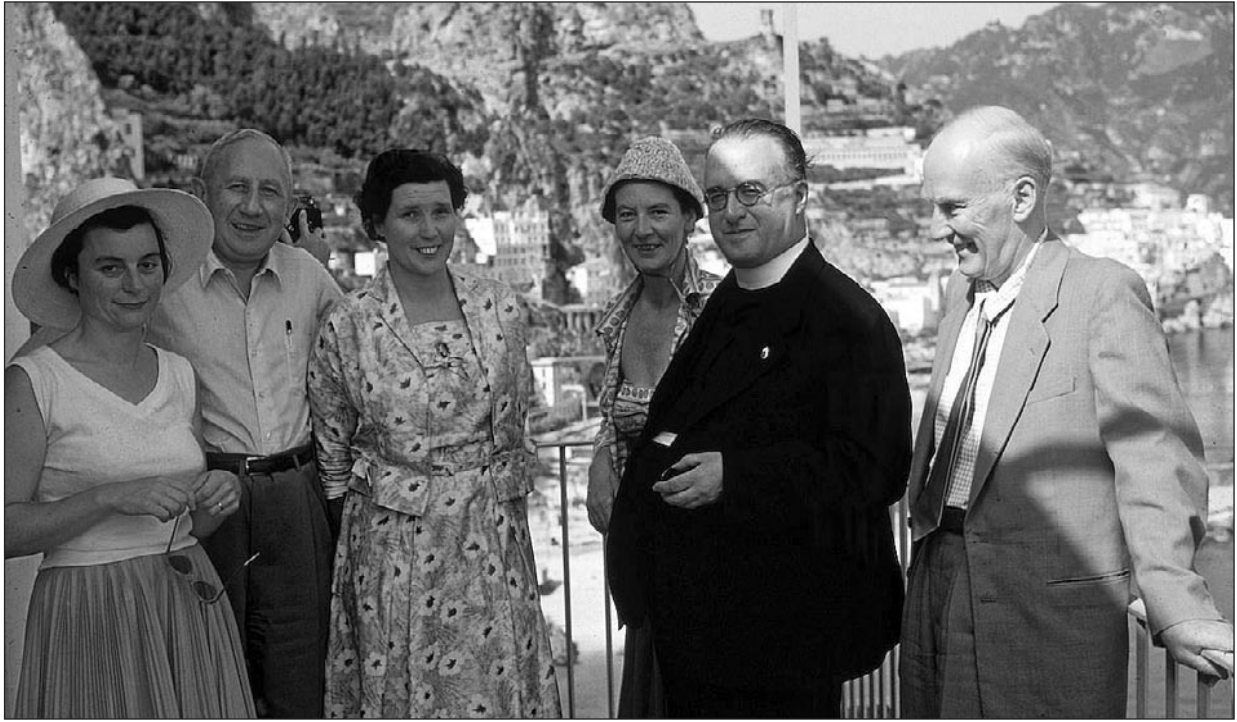

Fig. 5: Barbara Hoyle, Walter Baade, Muschi Baade, Adriane Fowler, Georges Lemaître, and Jan Oort relaxing on the Italian coast at Amalfi in 1957 May following a small conference at the Vatican on stellar evolution. Fred and Barbara Hoyle gave Lemaître a lift back to Belgium. Lemaître consumed vast pasta lunches and much wine, which he slept off in the back of the car. Barbara Hoyle was impressed that in Italy a Canon of the Roman Church was never billed for his restaurant meals. (Master and Fellows of St John's College, Cambridge)

In arriving at his sensational conclusions, Hubble had employed the de Sitter model, so the debate on the merits of that model and of Einstein's was revived. At the Royal Astronomical Society on 1930 January 10, de Sitter gave an informal presentation on the recession velocities of nebulae. He noted that the real Universe had sufficient matter for an Einstein universe, and enough motion for the de Sitter universe, which suggested the need for an intermediate solution.[36] Eddington agreed wholeheartedly and set to work on this task, together with his doctoral student George Cunliffe McVittie (1904–1988).

When Lemaître spotted the summary of the January RAS meeting in *The Observatory* magazine he wrote to Eddington drawing attention to his 1927 paper and its solution. Eddington was horrified when he realized that he had no recollection of having received the copy that his former student claimed had been dispatched. McVittie later recalled that a shamefaced Eddington had shown him the letter, exclaiming that Lemaître had already found the intermediate solution.[57]

Eddington took immediate action to publicize Lemaître's breakthrough. He added a final paragraph to a review of Silberstein's book that he had written for *Nature*:[58]

> Three years ago a very substantial advance in this subject was made by Abbé G. Lemaître … [which]

renders obsolete the contest between Einstein's and de Sitter's cosmogonies. We can now prove that Einstein's universe is unstable.[59]

On 1930 March 19 Eddington sent a copy of Lemaître's 1927 paper to de Sitter in Leiden, adding on the title page that 'This seems a complete answer to the problem we were discussing.' Four weeks later, on April 17, de Sitter wrote to Shapley about the new theory 'which must be somewhere near the truth'.

In May Eddington heaped more praise on 'Lemaître's brilliant solution' that allowed 'an infinite variety of spherical worlds that are not in equilibrium'. He seems to have had no doubt that Lemaître had discovered the expanding Universe.[60]

Lemaître had shown that it might have started as a static Einstein universe with a cosmological constant and remained in such a state for an indefinite period of time, until its instability became manifest. It then began expanding in a way that grew over time and will continue indefinitely.

### 8.1. Hubble gets angry with de Sitter

Throughout the summer of 1930, de Sitter and Eddington were promoting Lemaître's work and referencing it in several publications. On May 26 de Sitter produced a meticulously researched analysis of the data on the measured velocities and estimated distances of nebulae, including a few with large velocities.[61] He noted a strong correlation between radial velocity and distance for 32 galaxies.

De Sitter's paper is highly significant because of the depth of its mathematical analysis, which had far-reaching consequences. He had extended the velocity–distance relationship out to the point where it was plainly possible to discriminate between different cosmological models. The remarkable result he found is that neither Einstein's model nor his own could be true. These static models, he concluded, 'cannot represent the observed facts'.

A coda at the end of his paper adds that Eddington had informed him of Lemaître's paper 'only a few weeks ago', and that he hoped to return to 'this ingenious solution in a separate communication'.

When de Sitter's paper landed on Hubble's desk in 1930 August it sparked a furious reaction. Hubble rebuked de Sitter for the casual manner in which the paper was presented, and its failing to acknowledge the Mount Wilson contribution to the velocity–distance relation. And he reproached de Sitter for lifting 'data that has appeared in Annual Reports, etc' before Mount Wilson had been able to work on the cosmological implications of their own observations.[62]

Fortunately, the two made peace within a year. De Sitter's promised 'separate communication' came in 1930 June.[63] It was a lengthy discussion of Lemaître's solution, to which the latter replied in July.[64] This pair of papers appear to be the first to refer to 'the expanding universe' in their titles.

### 8.2. *Redaction by the RAS*
The Council of the Royal Astronomical Society sought Lemaître's permission to publish a translation in their *Monthly Notices*, 'in the honour of giving your paper a greater publicity amongst English speaking scientists'. Lemaître personally provided the translation. In 1931 February he received the formal letter of invitation from RAS Secretary William Marshall Smart (1889–1975), a powerful figure in the British astronomical establishment. Smart stated that the translation should be of paragraphs 1–72 only.[65] The suppressed paragraph 73 included Lemaître's determination of the coefficient of expansion (i.e. the Hubble constant).

We do not know what motivated Smart to censor the translation; perhaps he did not wish to see the Society upset Hubble. Whatever the case, Lemaître already knew of Smart from his year in Cambridge. Therefore, he dutifully implemented the request: the discussion and use of radial velocities of galaxies and their distances was redacted on the grounds that the data to hand in 1927 had been superseded.

An ironic consequence of Smart's instruction was that the target readership of 'English speaking scientists' thus remained ignorant of Lemaître's pioneering fusion of observation and theory, announced two years *before* Hubble at Mount Wilson handed down confirmation of the velocity–distance relation.

Much later, Lemaître came to regret caving in so submissively. By 1950 he felt the need to correct 'the history of this science competition', and he wished there to be no doubt that the motivation behind his great paper of 1927 was to explain the radial velocities of nebulae as a natural event 'in a universe with constant mass and increasing radius'.[66]

Note that Hubble never claimed to have discovered the expanding Universe; in fact, he probably never believed in such a scenario.[67] Furthermore, it would have been impossible for him to have discovered the expansion purely by observation.

## 9. The cautious cosmic celebrity
Thanks to the tireless campaigns of Eddington and de Sitter, the Belgian priest effortlessly rose to the status of cosmic celebrity. The cosmologists were content that they now had a theory that seemed to explain the present, and even predict the future of the Universe. There was a puzzling issue about the beginning, and we shall come to that shortly, but after Lemaître's translated paper appeared in 1931 March the expanding Universe became the accepted cosmological model. Lemaître, however, was conspicuously uncomfortable with the prospect of basking in glory. Historians are divided on the causes of Lemaître's reticence: had he become bored with cosmology; or doubtful as to the validity of his thinking; or shy of attention; or perturbed by Hubble's prickly state of mind? Or was he troubled by something else? We do not and cannot know.

In my view, one likely trigger was his devout commitment to traditional Catholic teaching on the sins of pride and envy, as well as the several New Testament parables on humility. Another factor to consider is that, from the late nineteenth century, there was a significant anti-science and anti-evolution movement in conservative circles in Rome. That would have made him reluctant to proclaim his revolutionary hypothesis of an infinite expanding universe.

With the passing of Cardinal Mercier on 1926 January 23, Lemaître lost his greatest supporter in the Church. Mercier had been a progressive, liberal theologian, who had defended scholars at Louvain from accusations of modernism and who sought to reconcile traditional Catholic philosophy with rapidly developing scientific knowledge. Anti-science sentiment only began to lose traction from 1950 after Pius XII, an enthusiast for modern science, published a papal encyclical saying that the study of evolutionary biology did not conflict with the teachings of the Church.[68]

### 9.1. *Popularizing the expanding Universe*
In 1931 March, Humason and Hubble submitted their monumental paper on the velocity–distance relation for publication.[69] This masterwork added 40 new velocities, and stretched the distance limit out to 32 Mpc, a spectacular eighteen times farther than the 1929 paper. The form of the relation (a straight line) remained the same apart from a small revision of the distance scale. They rounded their estimate of the Hubble constant to 560

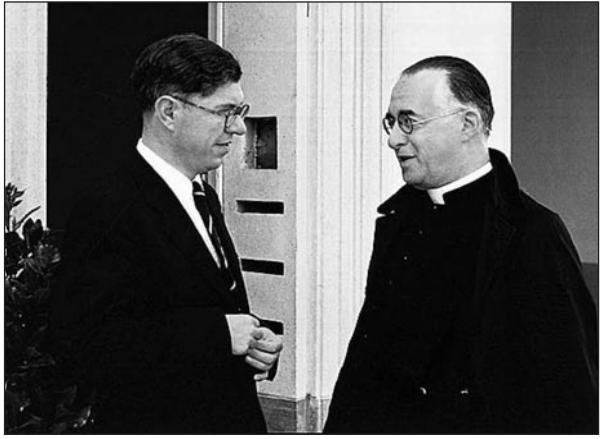

*Fig 6: Fred Hoyle, father of the Steady State theory, talks with Georges Lemaître, father of the Big Bang, in the late 1950s. (Master and Fellows of St John's College, Cambridge)*

km/s/Mpc, some 10 per cent below Lemaître's estimate of 1927. This paper established the observational reality of the velocity–distance relation but the search continued for the correct interpretation of the phenomenon.

On 1931 January 5, Eddington delivered an Address to the Mathematical Association, in his capacity as their President. Some three hundred teachers of mathematics in secondary schools gathered at the London Day Training College, established in 1902 by the great socialist reformer and economist Sidney Webb (1859–1947). Eddington's lecture was the first occasion on which he entertained a public audience with the news that 'we have recently learnt, mainly through the work of Prof. Lemaître, that … spherical space is expanding rather rapidly'.

To add to the fun, Eddington introduced a thought experiment through which he traced the state of the expanding world to the time when all of the matter and energy had the maximum amount of organization, a state of minimum energy. He continued, 'We would have come to an abrupt end of spacetime – only we generally call it the beginning.'

Eddington and many others felt that the question of a 'beginning' lay outside the realm of scientific enquiry, and Eddington certainly disliked it philosophically: 'the notion of a beginning of the present order of Nature is repugnant to me'.[70] He felt that way because the nature of our world is 'far from a fortuitous concourse of atoms'. For Eddington the idea of a 'beginning' was a confusion between pure physics and the theology of creation.

The text of the lecture published in *Nature* has a slightly shorter introduction than the one he gave to the school-teachers, in that an apologia is omitted:[71]

> I am afraid that the title of my address is not quite explicit. It lacks that precision of statement which ought to characterise the utterances of a mathematician. I have undertaken to speak to you about the End of the World, but I have not told you *which end*.

The address that followed was a deep examination of the concept of 'beginning' rather than 'end', with much talk of entropy and the heat death of the Universe. Eddington struck a thoroughly modern note in the lecture when he introduced Heisenberg's uncertainty principle, published in 1927, saying that it 'delivered the knock-out blow [because] it actually postulated a certain measure of indeterminacy or unpredictability of the future, *as a fundamental law of the universe*' (emphasis added).

He continued that this fundamental uncertainty removed any certainty about time in classical physics. In the new quantum world 'each passing moment brings into the world something new' that a mere mathematical extrapolation could not handle.

## 10. The fireworks universe

Eddington's former student strongly disagreed that the notion of a 'beginning' was repugnant. Lemaître sent a brief rebuttal to *Nature* in which he imagined that the beginning of spacetime can be considered as the disintegration of a single quantum, the Primeval Atom.[72] This note was not a serious scientific paper – he signed it off as a member of the public writing from a private address.

Lemaître's universal quantum had been dormant since 'before the beginning of space and time', which had only come into existence after the primordial entity had disintegrated because of its intrinsic instability. Lemaître outlined the process as the division of 'smaller and smaller and smaller atoms by a kind of super-radioactive process' due to the uncertainty principle. It was an extraordinary idea but put the early Universe in a quantum mechanical setting, only a handful of years after the foundation of modern quantum mechanics.

This line of thought, that radioactive processes were important in the early Universe, had been fermenting in his mind since his meeting with Millikan in 1926. By 1930, Lemaître had already reached another startling conclusion: 'One could admit that *the light* was the original state of matter and that all the matter … was formed by the process proposed by Millikan.'[73]

Millikan was not remotely the first scholar to question if the Universe had begun with light. When Lemaître was in the seminary studying the philosophy of Thomas Aquinas (*c.*1225–74), he would have encountered the works of Robert Grosseteste (*c.*1175–1253), who was for a time chancellor of Oxford University. A medieval scholastic, Grosseteste became greatly admired for his insistence on the use of mathematics in physics, his emphasis on empirical observation, and above all for his speculations about the nature of light.[74]

Grosseteste's original research into astronomy, mathematics, and physics had led him to anticipate modern cosmological ideas. His highly original treatise on light, *De Luce* (1225), postulated that light had preceded anything material, and that our world had materialized from pure energy exploding from a point source. It is credible that *De Luce* may have been one of the inspirational sources that led Lemaître to the Big Bang hypothesis.

### 10.1. *Cosmic rays are 'smoke and ashes'*

Lemaître's first opportunity to present a well-argued version of his joined-up thinking on quantum theory and cosmology came in 1931 October, at a meeting of the British Association convened by Herbert Dingle (1890–1978), who worked in astronomical spectroscopy at Imperial College. At the time Dingle had the highest regard for Eddington and Einstein. Like Eddington, he wrote for the public in a beautiful and clear literary style.[75] As early as 1922, he published *Relativity for All*,[76] and a useful textbook on modern astrophysics followed two years later.[77]

The meeting was a great public success, attracting a crowd of over two thousand, all eager to learn more about relativity, which was then still a highly esoteric subject. Dingle lined up Eddington, James Jeans (1877–1946), de Sitter, Millikan, and several others. Dingle had overlooked Lemaître, but this omission was corrected after an intervention by Eddington.

Lemaître delivered his talk following an opening address by Jeans, who hedged his bets, suspecting that the concept of an expanding universe might prove after all to be a false scent, with the truth lying in some other direction[78] Lemaître's response was forthright, transparent and unambiguous: 'The expansion of the universe is a matter of astronomical facts interpreted by the theory of relativity … I shall not discuss the legitimacy of this interpretation, as I do not know of any definite objection made against it.'[79]

From the observed rate of expansion, he arrived at a 'round numbers' age for the Universe of ten billion years, consistent with the continuing presence of radioactive uranium and thorium in the Earth's crust. Emphasizing that radioactive disintegration is a physical fact, and that cosmic rays are similar to the radiation emanating from radium, Lemaître suggested that cosmic rays must have been released by explosive radioactive disintegration of a primal atom at the onset of the Universe: cosmic rays were 'the ashes and smoke of bright but very rapid fireworks'. The Fireworks Universe no less!

De Sitter spoke next, brimming with confidence: 'There can be not the slightest doubt that Lemaître's theory is essentially true, and must be accepted as a very real and important step towards a better understanding of Nature.'[80]

Eddington agreed with de Sitter that the facts indicated a rapid expansion phase at the beginning of the Universe, and that an age of ten billion years should be acceptable to all, despite feeling that rapid expansion is 'so preposterous that we naturally hesitate before committing ourselves to it'.[81]

The British Association meeting offered all the features of a public entertainment: a huge audience, world-famous scientists using verbal communication rather than graphs and mathematics, and no in-depth nit-picking questioning of the speakers afterwards. Unsurprisingly, the wordy account of the talks published by *Nature* had no significant impact on the theoretical physics community.

### 10.2. *'Speculation run mad'*

Lemaître chose to publish a full version of his fiery primeval-atom model in French in *Revue des Questiones Scientifiques*.[82] This semi-popular journal had been set up in 1877 by the Scientific Society of Brussels, a Catholic organization that sought a rapprochement between the Church and modern science at a period of unrelenting anti-clericalism in France and Belgium.[83]

The 1931 Fireworks Universe paper gradually diffused into the minds of cosmologists and they were not impressed. Eddington never accepted the Fireworks Universe or indeed any suggestion that the expanding Universe had its origin in a Big Bang. Plaskett mounted a withering attack on Lemaître's 'wonder world', declaring the fireworks theory 'an example of speculation run mad, without a shred of evidence'.[84] The English mathematician Ernest William Barnes (1874–1953), a former Fellow of Trinity College, Cambridge, who might have remembered young Arthur Eddington as an outstanding undergraduate, remarked in a book on science and religion that 'many cosmogonists have yet to be persuaded by … a brilliantly clever *jeu d'esprit* – a flight of fancy of Lemaître's vivid imagination.[85] The American physicist Richard Chace Tolman (1881–1948) of Caltech cautioned against 'the evils of autistic and wish-fulfilling thinking' in cosmology.[86]

*Fig 7: L'Hypothèse de L'Atome Primitif, published in 1946, was a collection of five lectures delivered by Lemaître between 1929 and 1945. The book had only a small circulation, and surviving copies in good condition are very rare. (Simon Mitton)*

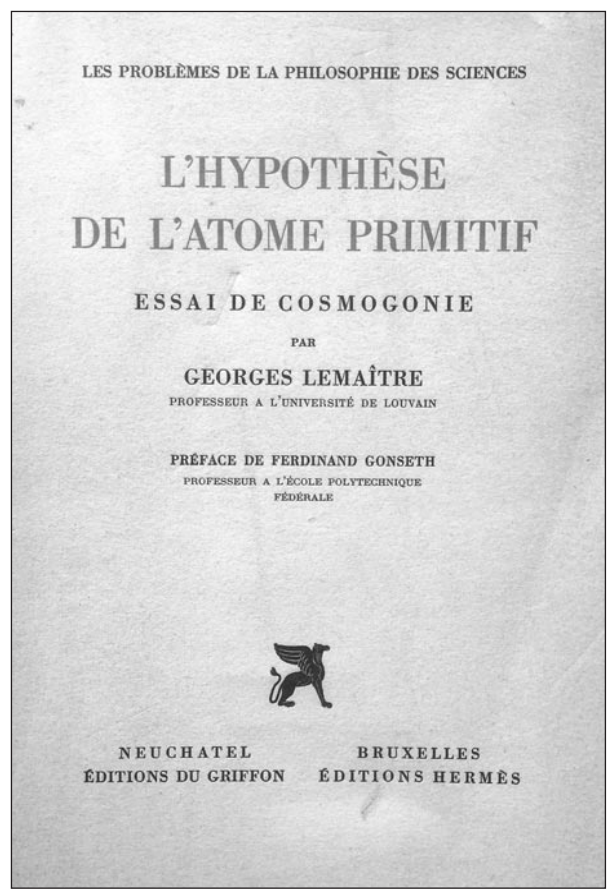

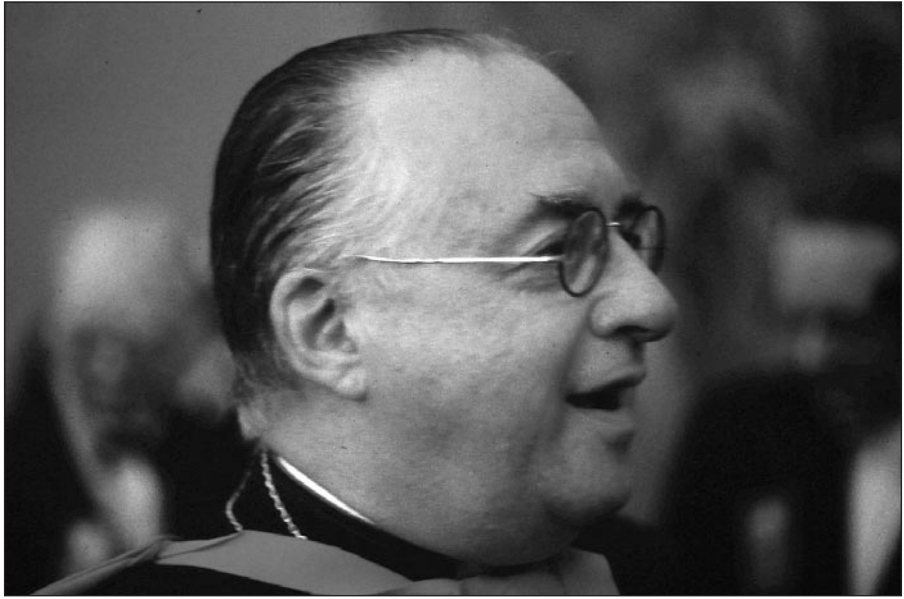

In the face of such criticism, Lemaître wrote to Eddington that he was taking comfort from support by Einstein, who had become an enthusiast for the fireworks model.[87] On 1933 November 20, Lemaître addressed the National Academy of Sciences on the evolution of the expanding Universe. The key point he made was to associate a negative pressure with the energy density of the vacuum, and to align the cosmological constant with that negative pressure. It was no longer a fudge factor – it had a physical meaning.[88]

Support of a different kind came from Edward Arthur Milne (1896–1950), professor of mathematics at Oxford. In the 1930s Milne opened an independent path. He sought a different mechanism of expansion and cosmic evolution that blended Newton's mechanics with Einstein's special theory of relativity. His approach did not include gravitation because he rejected general relativity.

By 1932 May he had his very own Milne universe, which began at time zero in a very small space, into which the galaxies (nebulae) were originally crammed into a very small volume. The galaxies were in a bubble that expanded at the speed of light into empty space, each galaxy coasting along at an arbitrary velocity.[89]

Milne had set off a great philosophical debate among the theoreticians that lasted until 1950. Dozens of papers appeared as a result of Milne's unorthodox approach and his challenge to the foundations of physics. Two consequences of the community's change of focus were the almost complete neglect of Lemaître's Big Bang solution and his own disengagement from cosmology.

### 10.3. *Publication of the Primeval Atom*

From 1935 until his death, Lemaître conducted research on cosmic rays. His taste for intellectually challenging mathematical puzzles led him to study the complex motion of cosmic-ray particles in the Earth's magnetic field. Because the equations governing such motion cannot be solved analytically, Lemaître became a pioneer of using computers for the numerical work: Louvain purchased a Burroughs E101 desk computer in 1958, and Lemaître was one of its first European users. From 1933 to 1945 he published a steady stream of theoretical papers on the physics of cosmic rays.

A flickering flame of interest in the expanding Universe was kept alight by his acceptance of many invitations to give set-piece lectures at national and international meetings. The published reports of such gatherings, as well as coverage in popular science journals, brought his cosmological ideas to a French-speaking public. In 1945 September he fulfilled an invitation to speak in Fribourg, at the annual meeting of the Swiss Society of Natural Sciences, on the subject of the Primeval Atom hypothesis.[90]

One positive outcome of that engagement is that the Neuchatel educational publisher *Editions du Griffon* produced a volume in French of three of his lectures on the grandeur of space (1929), the expansion of the Universe (1931), and the evolution of the Universe (1933). To these they added one on hypotheses of cosmogony (1945) and the talk given in Fribourg (Figure 7). The resulting popular book was translated into Spanish (1946) and English (1950).

The book's preface is by the Swiss philosopher and mathematician Ferdinand Gonseth (1890–1975), who highlighted two arguments for the primeval-atom hypothesis for which compelling evidence was present.

These were: the presence of radioactive elements with half-lives comparable to the Hubble time, and the law of entropy as applied to the decay of the primeval atom.

Lemaître's philosophy was that of a realist, unlike most cosmologists at the time: he was the first to be convinced that expansion was real and that its cause must be sought. For a general talk to the Catholic Institute in Paris in about 1950 he used the title 'The universe is not beyond human possibilities'.[91]

## 11. Reigniting the Big Bang

Interest in models of the expanding Universe was revived by the nuclear physicist George Gamow (1904–1968) in 1948. From 1923 he studied mathematics and physics in Petrograd (later Leningrad), where he attended Friedman's lectures on general relativity. Unfortunately, his intention to work in relativistic cosmology was stymied by fate, so Gamow turned his mind to the physics of the nucleus in his early career.

After moving to George Washington University in 1935, Gamow continued in nuclear physics, switching to theoretical astrophysics from 1938. During the mid-1940s he changed fields again, taking up relativistic cosmology by rising to the challenge of accounting for the origin of the chemical elements in the expanding Universe.[92]

Gamow asked his talented student Ralph Asher Alpher (1921–2007) to investigate whether the formation of the heavier elements could have taken place in a hot primordial gas. Alpher undertook the daunting challenge of solving numerically the equations governing element synthesis. He and Gamow concluded that the conditions for nuclear synthesis had lasted for only 300 seconds: the chemical elements had been forged in a hot Big Bang.[93]

During his preparation for doctoral research, Alpher had made a systematic study of all the papers then published on cosmology, including Lemaître's recent book. However, Gamow and his co-workers had no reason to connect Lemaître's speculative work in cosmology with their conclusions on nuclear processes in the early Universe.[94]

From 1948 to 1953 Gamow, Alpher, and Robert Herman (1914–1997) continued their work on the conditions of temperature and density in the evolving early Universe during the fleeting era of element building.[95] Alpher, aware that the intense radiation during the hot phase of the beginning of the Universe would leave a fossil signature, calculated that it would have cooled to 5 K today. The cosmic microwave background is that fossil, and its actual temperature is 2.73 K.

### 11.1. Hoyle's Steady State theory
In the late 1940s, cosmology was still a niche subject, pursued by some two dozen applied mathematicians, physicists, and astronomers. There was no institutional framework supporting the subject, and its practitioners were not yet organized as research groups.

While Gamow was reformulating the Big Bang theory to include nuclear physics there appeared a 'new cosmology' in Cambridge, England. Fred Hoyle, Hermann Bondi (1919–2005), and Thomas Gold (1920–2004) responded to Gamow's hot Big Bang model with a diametrically opposed proposition that soon came to be known as the Steady State theory. The essence of their proposal was that the Universe had always existed, had the same properties and appearance everywhere in space and time, and that the continuous creation of new matter filled the void left by expansion.

Hoyle in particular disliked the notion of an initial cause beyond the realms of science, which was required by the Big Bang, and tended to dismiss any theories 'requiring a state of knowledge for which we have no evidence'.[96] Hoyle himself coined the term Big Bang in a BBC radio programme broadcast on 1949 March 28 to describe the model of the Universe as expanding from a primordial condition of enormous density and temperature. It took two decades for the expression Big Bang to catch on; Hoyle never used it 'pejoratively' as Gamow falsely claimed.[97]

The Steady State theory failed to attract significant support from the professionals despite the adulation with which it was received by the general public. Bondi commented in 1952 that 'Lemaître's model ... seems to be the best relativistic cosmology can offer'.[98] Thus it was that from the early 1950s the Big Bang theory steadily gained ground, and yet Lemaître faded into the background, almost certainly because he had ceased to promote his brilliant idea.

On 1998 September 1, *The Astronomical Journal* published a spectacular paper announcing the discovery of the accelerating Universe. An international effort to measure the distances of sixteen high-redshift supernovae had established that they are on average '10–15 per cent farther than expected in a low mass universe without a cosmological constant'.[99]

Cosmologists ascribed this acceleration to unseen dark energy pervading the universe and having the same effect as a negative pressure – the cosmological constant. The excitement at the time was palpable, and I vividly remember my cosmology colleagues in Cambridge exclaiming that 'Lemaître's cosmological constant is inflating the universe'.

Overnight it seemed that a huge catch-up industry had sprung up, with a younger generation of cosmologists revelling in the discovery of Lemaître's papers. Over the past two decades the historian Dominique Lambert in Louvain has provided us with a truly exceptional biography of Lemaître, while Helge Kragh of the University of Aarhus has meticulously documented the contributions of many observers and cosmologists to the foundations of the concordant cosmology that we have today. From these new assessments, we can see that Georges Lemaître, the father of the Big Bang, was arguably the greatest cosmologist of his generation.

## Acknowledgements

I am most grateful to my colleagues Rodney Holder, Michael Hoskin, Cormac O'Raifeartaigh, Ian Ridpath, and Michael Robson, and to my wife Jacqueline Mitton, for great advice and support.

**The author**

Simon Mitton is an emeritus academic at the University of Cambridge. During his doctoral training at the Mullard Radio Astronomy Observatory he worked closely with Martin Ryle on mapping radio galaxies and quasars. For several years he was departmental administrator at the Institute of Astronomy, following which he had a long career as the executive director of science publishing at Cambridge University Press. He returned to research in 2001, taking up a new career in the history of astronomy in the twentieth century, on which he has published several books and papers. He is a Life Fellow of St Edmund's College, Cambridge, where Georges Lemaître resided 1923–1924.